\documentclass[a4paper,showkeys,floatfix,aps,prl,reprint,groupedaddress,singlecolumn]{revtex4-1}
\usepackage{caption}
\usepackage{amsmath,amssymb}
\usepackage{graphics,graphicx}
\usepackage{dcolumn,bm}
\usepackage{psfrag}

\begin{document}

\title{Maximizing the Strength of Fiber Bundles under Uniform Loading}



\author{Soumyajyoti Biswas${}^{1,2,}$}
\email[]{sbiswas@imsc.res.in}
\author{Parongama Sen${}^{3,}$}
\email[]{psphy@caluniv.ac.in}
\affiliation{
${}^1$ The Institute of Mathematical Sciences, Taramani, Chennai-600113, India.\\
${}^2$ Earthquake Research Institute, University of Tokyo, 1-1-1 Yayoi, Bunkyo-ku, Tokyo-1130032, Japan.\\
${}^3$ Department of Physics, University of Calcutta, 92 Acharya Prafulla Chandra Road, Kolkata-700009, India.
}


\begin{abstract}
\noindent  The collective strength of a system of fibers, each having a failure threshold drawn randomly from a 
distribution, indicates the maximum load carrying capacity of different disordered systems ranging from disordered solids,
power-grid networks, to traffic in a parallel system of roads. In many of the cases where the redistribution of load following a local failure 
can be  controlled, it is a natural requirement to find the most efficient redistribution scheme, i.e., following which  system can
carry the maximum load. We address the question here and find that the answer depends on the mode of loading. We
analytically find the maximum strength and corresponding redistribution schemes for sudden and quasi static loading.
The associated phase transition from partial to total failure by increasing the load has been studied. 
The universality class is found to be dependent on the redistribution mechanism.  
\end{abstract}


\maketitle

\noindent The response of an ensemble of elements having random failure thresholds plays
a crucial role in the breakdown properties of heterogeneous systems \cite{books}. Such systems can be,
 for example, 
disordered solids under stress, power-grid networks carrying currents, roads carrying car traffic, or redundant
computer circuitry etc. 
The failure properties of these systems are  strongly dependent upon the load redistribution mechanism
following a local breakdown. While in some cases (e.g. solids under stress) such mechanisms are 
properties inherent to the system, in many other cases (e.g. power-grids \cite{power-grid}, traffic controls \cite{traffic}) it is a 
matter of design that can be optimized so as to achieve the most robust configuration (see e.g. \cite{robustness}).  Such robustness
properties of networks connecting elements with varying failure thresholds  under targeted or random attacks 
have received substantial attention \cite{attack1, attack2}.

In this Letter,  we address the question of maximizing the strength of a disordered system by finding the
most effective redistribution scheme analytically using the fiber bundle model as a generic example. This model was introduced
\cite{first}, indeed, to estimate the strength of cotton in textile engineering. 
Since then, it has been extensively studied in the context of distribution of failure strength and failure time of disordered materials under
tensile loading or twist by viewing fibers as elements of the disordered solids having a finite failure threshold (or even a finite
lifetime dependent of loading) \cite{daniels,phoenix1,phoenix2,phoenix3,phoenix4,phoenix5,daniels2,leath,zhou,phoenix6,curtin,phoenix7,phoenix8,phoenix9,phoenix10} 
(see \cite{rmp1} for a review). 

Conventionally, the fiber bundle model is viewed as a set of parallel fibers, having failure thresholds randomly drawn from a 
distribution (say, uniform in $[0:1]$), clamped between two horizontal plates. When the bottom plate
is loaded, some of the weak fibers break, and their load is redistributed among the surviving fibers, which may in turn break or
survive depending on their failure thresholds. The load transfer rule, generally a function of distance from the broken fiber, 
plays a vital role (see e.g., \cite{arne, new}). 
 While substantial attention has been concentrated
in revealing the transition properties and related dynamics at and near the failure point, much less is known about  tuning the redistribution scheme to maximize the strength of the system.   
In many  applications, as mentioned above, this is a matter of design and hence remains an important
question.

We consider the system as simply a set of interconnected elements and the 
load transfer is heterogeneous and dynamic in the sense that it may depend on the threshold 
and also on the instantaneous values of the loads on different elements.
Over the years, examples of time dependent failure, which could sometimes be translated to 
some sort of threshold dependence,
were studied (see for example \cite{phoenix2, curtin,creep}). However, it is hard to make explicit comparisons here.

Before entering into the details of the heterogeneous scheme, we note that the loading 
process can be broadly divided in two ways: sudden loading and quasi-static or gradual loading.
We also note that the external loading is always uniform although redistributions can lead to
inhomogeneities in load carried by individual fibers.

In the case of {\it sudden loading}, a load  $W$ is
applied uniformly on the initially intact and unloaded system in a step function, consequently 
the load per fiber initially is $\sigma=W/N$ ($N$ being system size, i.e. number of fibers). 
The redistribution schemes then follow the initial breaking, leading 
to a stable state which is either partially damaged or has completely failed. 
Irrespective of the redistribution scheme followed, an upper bound
of failure threshold (say, $\sigma_m$) can readily be estimated for this case.  
On application of stress $\sigma_m$ to each fiber, those having thresholds $\leq \sigma_m$
will immediately break.
The best possible scenario is that no more fibers break and they share the total 
load and carry load to their highest capacity.
The average threshold of the remaining fibers is $(1+\sigma_m)/2$ and the fraction of surviving fibers is $(1-\sigma_m)$
(threshold distribution uniform in $[0:1]$). The condition of maximum strength becomes  $(1-\sigma_m)(1+\sigma_m)/2 = \sigma_m$
i.e., $\sigma_m=\sqrt{2}-1$. For arbitrary threshold distribution, the condition is $\int\limits_{\sigma_m}^{\infty}fp(f)df=\sigma_m$,
where $p(f)$ is a threshold distribution function, i.e. total remaining threshold should be equal to
total load  (see Supplemental Material). 
This  limiting value is obtained without assuming any particular load redistribution scheme 
and it is worthwhile to find out  at least one scheme which can lead to 
this limiting behavior. 
 
The uniform load redistribution is a homogeneous scheme which can give 
a loading capacity equal to $1/4$ \cite{rmp1}, much less than the maximum value derived above. 
Hence we use a heterogeneous scheme  to explore the possibility 
of obtaining a failure threshold larger than $1/4$.

 Particularly, we 
have considered the case where the load received by the  $i$-th fiber in a given redistribution 
process is proportional to $(f_i-\sigma_i)^b$, where $\sigma_i$ and $f_i$ are respectively the
load and failure threshold of the $i$-th fiber. The limit $b=0$ is the well studied \cite{rmp1} uniform load redistribution.
We have numerically obtained the critical loads for different $b$ values, which also depend on 
whether the load is applied suddenly or quasi-statically. We could analytically calculate the
limiting cases of critical load and have found that the most robust condition (highest value of critical load) 
arises for $b=1$
in sudden loading and $b\to \infty$ for quasi-static loading (effectively for $b\approx 10$ or above).

 Furthermore, from the point of view of phase transition between partial and complete failures, we 
have found that for $0\le b<1$, the order 
parameter exponent value remains the same. But for $b\ge 1$ it is different, showing a different
universality class than that of the uniform load sharing one. The dynamics, and hence the system size dependence 
of the relaxation time near the critical point depends strongly on the loading mechanism.

\begin{figure}[t]
\centering 
\includegraphics[width=10cm]{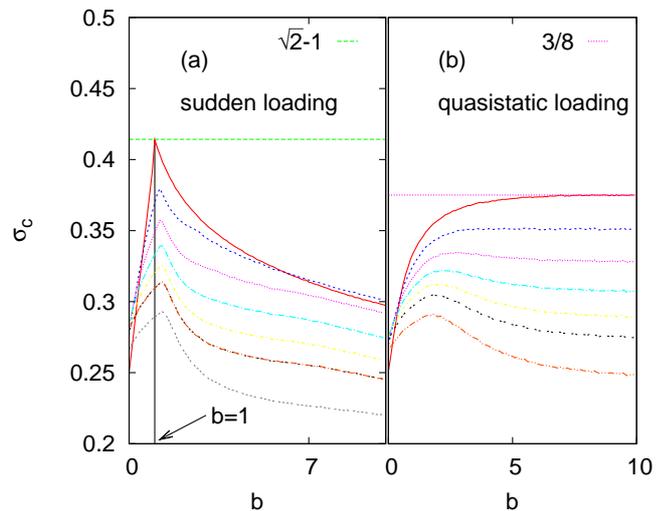}
   \caption{(Color online) The phase diagram in the $b-\sigma_c$ plane is shown for (a) sudden and (b) quasistatic loading for various fractional errors in the
knowledge of the threshold values of the individual fibers (curves are from top to below for $e=0.0, 0.1, 0.2, 0.3, 0.4, 0.5, 0.75$). The upper bounds for both cases are shown
which are reached for $b=1$ (a) and  $b \to \infty$ (b).}
\label{ph_d_beta}
\end{figure}

A natural expectation for a better scheme is to assume that the stronger fibers should get a higher share. 
However, the {\it effectively} stronger fibers are
those which have higher difference between the failure threshold and the load it presently carries. Hence, we consider that the load received by
the $i$-th fiber having stress $\sigma_i$ and failure threshold $f_i$ is 
 $x_i=A(f_i-\sigma_i)^{b}w_e$, where $w_e$ is the excess load to be redistributed in a given step and $b$ is a parameter. 
In general, the redistribution can be a many step process, but let us consider the first step only, such that
$\sigma_i=\sigma$ and $w_e=\sigma^2N$ (since in the first step $\sigma N$ number of fibers will break each having load $\sigma$). 
The normalization (or, load conservation) condition reads 
\begin{equation}
\sum\limits_{i\in surviving fibers}\int\limits_{\sigma}^1\left[A(f_i-\sigma)^b\sigma^2N+\sigma \right]P(f_i)df_i=\sigma N, 
\end{equation}
with $\sigma N$ being the applied load and
$P(f)\propto 1/(1-\sigma)$ (this is essentially the normalized distribution of thresholds of remaining fibers),
 giving $A=(b+1)/(N(1-\sigma)^{b+1})$ (see Supplemental Material for calculation).
Hence the load received by the $i$-th fiber in the first redistribution step, having failure threshold $f_i$ following 
the sudden application of load per fiber value $\sigma$ is
\begin{equation}
  x(f_i)=\frac{\sigma^2(b+1)}{(1-\sigma)^{b+1}}(f_i-\sigma)^b.
\end{equation}
Now, the redistribution will occur only once (no recursive dynamics) if after the first redistribution all fibers have force lower
than their respective thresholds i.e.,  $x(f_i)+\sigma<f_i$ for all $i\in surviving fibers$. 
(i) For $b=1$, this inequality leads to
\begin{equation}
  f_i\left[\frac{2\sigma^2}{(1-\sigma)^2}-1\right] > \sigma\left[\frac{2\sigma^2}{(1-\sigma)^2}-1\right],
\end{equation}
which is generally satisfied (since for surviving fibers $f_i>\sigma$) upto a critical $\sigma_c$ where $2\sigma_c^2/(1-\sigma_c)^2=1$
or, $\sigma_c=\sqrt{2}-1$. But this is  the maximum possible sustainable load per fiber value, implying
 for $b=1$, the maximum strength is achieved and that is  a single step redistribution (see Fig. \ref{ph_d_beta}); 
(ii) For $b\ge 1$, the condition for recursive dynamics leads to a lower bound for failure strength, given by
$\sigma_l=\frac{\sqrt{1+b}-1}{b}$ (see Supplemental Material for derivation).

\begin{figure}[t]
\centering 
\includegraphics[width=8cm]{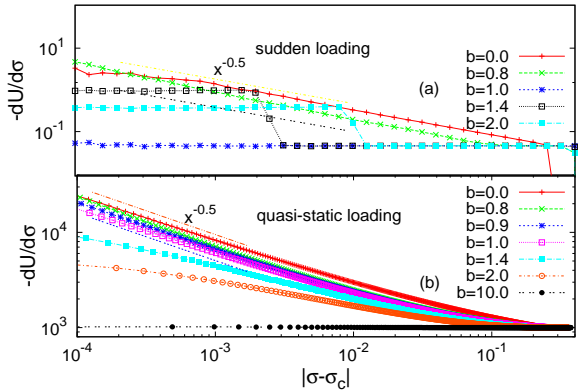}
\caption{(Color online) The variations of the derivative of order
parameter are shown for different b values for
  (a) sudden  and (b) quasi-static loading. The
  quantity is inverse of $1-\zeta$  (see Eq. (\ref{op_br})), which is also
     used as an alternative order parameter. These indicate that the order parameter exponent value
       $\beta = 1/2$, for $b <  1$. For $b \geq  1$, $\beta^\prime = 0$, i.e. the curves
               gradually become horizontal. } 
\label{op_derri}
\end{figure}

\begin{figure}[t]
\centering 
\includegraphics[width=8cm]{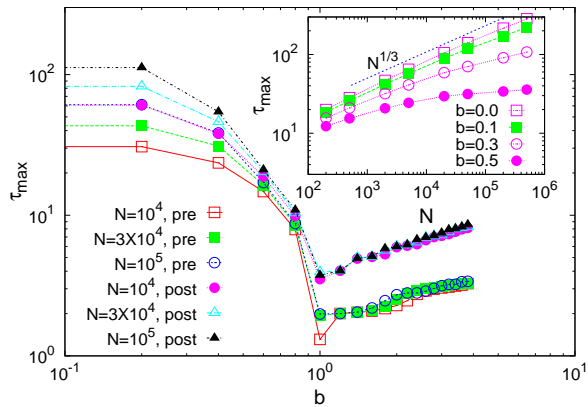}
   \caption{(Color online) The main panel shows the variations of maximum relaxation times (pre-critical and post-critical) with $b$ for different
system sizes. As $b\to 0$, the $N^{1/3}$ behavior is maintained (see inset). But as $b$ is increased, the power-law behavior 
is lost. For $b\ge 1$, system size dependence is almost non-existent. These data are for sudden loading. For gradual loading similar
behavior is seen, as the two versions are equivalent as far as the dynamics on or around critical point is concerned.}
\label{relax_time}
\end{figure}
The exponent values of the failure transition can be estimated for different $b$ values. Particularly, the order parameter 
exponent value can be analytically found for $b=1$ and is known for $b=0$. The well studied case of $b=0$ gives the
order parameter variation as $O=U(\sigma)-U(\sigma_c)=C(\sigma_c-\sigma)^{\beta}$, with $\beta=1/2$ \cite{rmp1}, where $U(\sigma)$
is the fraction of surviving fibers under a load per fiber value $\sigma$. This definition is,
however, difficult to follow in the numerical measures where the quantity $U(\sigma_c)$ is not known exactly. A different
 order parameter measure, called the branching ratio $\zeta$, was proposed in \cite{branching_ratio} (see also \cite{moreno}). It was defined as
$\zeta=1-n_0/\sum\limits_{t=0}^{t_{max}}n_t$, where $n_0$ is the number of failed fibers immediately following
a load increase (necessarily small amount) and the sum is the total number of fibers failed due to these initial
failures. Then the quantity $1-\zeta$ near the critical point follows $1-\zeta \sim (\sigma_c-\sigma)^{\beta^{\prime}}$.
The exponents $\beta$ and $\beta^{\prime}$ need not be equal in general. The quantity $\sum\limits_{t=0}^{t_{max}}n_t$
is essentially the change in the number of fibers following a small increase in the load, or in other words is
proportional to $|dU/d\sigma|$. But near the critical point 
\begin{equation}
\left|\frac{dU}{d\sigma}\right|^{-1}\sim 1-\zeta\sim (\sigma_c-\sigma)^{\beta-1}.
\label{op_br} 
\end{equation}
This leads to
the relation $\beta=1-\beta^{\prime}$. Incidentally, for $b=0$, $\beta=1/2$, giving $\beta=\beta^{\prime}$ in that case.
However, for $b=1$ the redistribution process has been shown to be single step and $U(\sigma)=1-\sigma$. Therefore,
$O=U(\sigma)-U(\sigma_c)=(\sigma_c-\sigma)$, giving $\beta=1$. On the other hand, since there is no avalanche in this case until the
failure point, $1-\zeta=1$ for $\sigma<\sigma_c$ and $0$ at $\sigma=\sigma_c$ (one may think this as a zero value of the exponent $\beta^{\prime}$).

Interestingly,  the exponent $\beta$  seems to have a value close to $0.5$ 
 for $b<1$
(see Fig. \ref{op_derri} (a); (b) shows the same for gradual loading).
For $b > 1$,  $\beta $ remains equal to $1$,  however, 
the  value of $|dU/d\sigma|$ shows a sudden change  at  $\sigma_l$,
where the redistribution process starts extending to more than one step. 

Another way to probe the critical behavior is to look at the relaxation time at (or very near to) the failure point. It is known that
at the critical point, the relaxation time scales as $\tau\sim N^{\alpha}$, with $\alpha=1/3$ for $b=0$ \cite{manna, new}. But 
for $b>0$ the power law scaling seems to be lost, with $\tau$ becoming almost independent of $N$ for $b>1$ (see Fig. \ref{relax_time}).
Here we have shown the behavior of both the pre-critical relaxation time (the relaxation time in the last stable configuration of the model)
and the post-critical relaxation time (the relaxation time in the first condition of complete failure). 

\begin{figure}[t]
\centering 
\includegraphics[height=6cm]{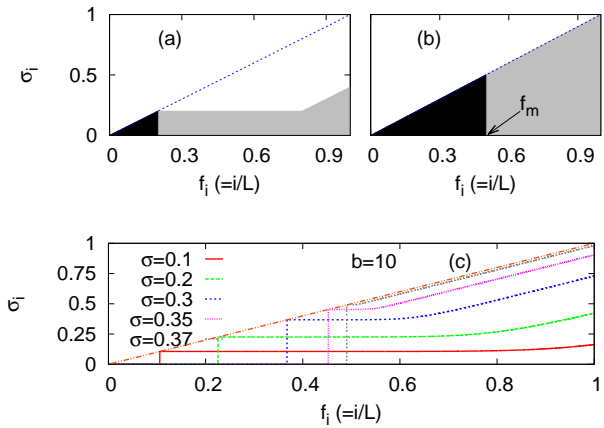}
   \caption{(Color online) (a) shows the schematic load profile in $b\to\infty$ limit 
when the gradual loading is done upto some threshold value. 
The fibers are arranged in increasing order of their threshold value.
 The black triangle on the
left are the broken fibers. The grey area denotes the load. There is a uniform external loading. The redistributed load of the
broken fiber is dumped on the right hand corner triangle, because for those fibers $f_i-\sigma_i$ were highest. (b) shows
the extreme limit, beyond which all fibers will break together. Simulation results for $b=10$ supporting the above picture
is shown at (c).}
\label{gradual_max}
\end{figure}
 Apart from  the sudden loading limit discussed so far, the other extreme is the quasi static loading. 
Here at every step the load is uniformly increased on all the surviving fibers until the weakest one breaks. The loading is then stopped
until the system goes to the next stable state by local failures and subsequent stress redistribution. 
After the relaxation to a stable state the system is further loaded to continue the dynamics (system relaxation assumed to be 
much faster than external loading rate).
 By observation it is seen that the 
failure threshold increases with $b$. For a very large $b$, any redistribution of the load will 
essentially be eaten-up by the fiber(s) with largest $(f_i-\sigma_i)$ value(s).  
Therefore, in the situation where the load is gradually increased upto the point where a fiber with strength $f$ has broken, the 
load profile will be like the one schematically depicted in  Fig. \ref{gradual_max} (a). The redistributed load 
will be piled up in the right hand corner
in the form of a triangle and there will be a uniform loading for the other fibers, which is due to the uniform load increase mechanism. 
Now, the extreme end of this process is the one depicted in  Fig. \ref{gradual_max} (b). Here $f_m=1/2$ and beyond this point all remaining
fiber has $(f_i-\sigma_i)$ values equal to each other. Hence, a further increase in the load will result in the complete collapse of the system.
This is, therefore, is the point where the system carries maximum load using gradual loading. The total load (per fiber) carried by the system at this point is
$\int\limits_{1/2}^{1}\sigma d\sigma=3/8$. Therefore, maximum failure threshold for gradual loading is $\sigma_c^{grad}=3/8$, for uniform threshold 
distribution in $[0:1]$, and is obtained in the limit $b\to\infty$. In practice, for $b\approx 10$ the maximum value is almost achieved.
Fig. \ref{gradual_max} (c) depicts the situation when the simulation is done for $b=10$.

The universality class in this case is determined  as before. The derivative of the order parameter diverges with $|\sigma_c-\sigma|$
with an exponent value $1/2$ as long as $b<1$ (see Fig. \ref{op_derri} upper panel). In this case, unlike the sudden-loading one, $b=1$ too seems to
belong to this universality class, at least within the numerical accuracies. For $b>1$, the divergence gradually flattens as the critical
point is approached, indicating a change in the universality class, as before. The behavior of the relaxation time remains the same (not shown) as in the
case of sudden loading.

  The maximum achievable strength in sudden loading is obtained for $b=1$  following the
load redistribution method proposed in this Letter; however this scheme requires the detailed knowledge of the failure thresholds of each element. In real
situations for controlled processes this knowledge can be  rather imprecise. The question is how effective these load redistribution strategies are
in those cases? Let us say the imprecisely known value of the threshold of a fiber $f_i^a$ can be anywhere between
$f_i(1-e)<f_i^a<f_i(1+e)$ with uniform probability, where $e$ is a constant $(< 1)$ and $f_i$ is the actual
threshold (our results so far are for $e=0$). We set the additional rule that the fibers 
already carrying loads equal to or higher than 
 their assumed thresholds will not get any further share of load from redistribution, 
which can happen when the assumed value is less than the actual value (due to this rule, even for $b=0$ the
critical load is higher than $1/4$, since, even for an imprecise knowledge,
generally the weaker fibers do not get load in redistribution). On the other hand,
if the assumed value is larger, it will break when the actual value 
is exceeded. 

 It turns out, even for $e>0$, this strategy can give better result than uniform
load redistribution for both sudden and quasistatic loading (see Fig. \ref{ph_d_beta}) with $\sigma_c$  
significantly higher for  a considerable range of value of $b>1$; the range decreasing with higher values of $e$.  
For the quasi static  loading case, the emergence of a shallow peak  in
  $\sigma_c$ occurring at a finite value of $b$ is to be noted. 
 These observations indicate the effectiveness of the present redistribution scheme in enhancing the  upper critical value of the load even with partial knowledge. 

While we have noted marked difference between the two extreme loading mechanism, there are instances for time dependent
fiber bundles \cite{phoenix7,phoenix10}, where these two limits can tend to converge. Detailed discussions on these are beyond the scope of the present manuscript and will be taken up elsewhere \cite{bs2}.

In conclusion, the threshold dependent load redistribution scheme proposed here can lead to 
significantly higher strength in the fiber bundle model, 
compared to the uniform load redistribution mechanism usually followed. For threshold distribution uniform in $[0:1]$, the maximum 
strength (load per fiber value prior to catastrophic failure)
 for sudden loading is found to be $\sqrt{2}-1$, which can be achieved for load redistribution proportional to
$(f_i-\sigma_i)^b$, with $b=1$.  For quasi static loading, the maximum strength is $3/8$ and is achieved in the limit $b\to\infty$.
While for sudden loading the maximum achievable strength is obtained using this scheme, for quasistatic loading whether
 there exists a more efficient redistribution scheme that
can lead to a higher value of maximum strength, remains an open question.
We also find that the universality class is different from the homogeneous loading case
for $b>1$ for both the loading schemes. 
The results remain qualitatively valid for other threshold distributions (Weibull, Gumbel, Gaussian) and
are also useful when the individual thresholds are known only approximately. 
The precise bounds obtained here and the corresponding load redistribution mechanisms can help in maximizing the strengths
of complex interconnected systems with an ensemble of elements each having finite failure threshold. Of course, this opens up the challenging question of practical
implementation, which will depend on the specific details of the system considered.

\acknowledgments PS acknowledges financial support from CSIR, Govt. of India
 and hospitality of The Institute of Mathematical Sciences, where majority of the work was done.

\begin{center}
{\large {\bf Maximizing the strength of fiber bundles under uniform loading: supplemental material}}\\
\end{center}

\section{Derivation of lower bound of critical load for sudden loading}
One can  show that for $b\ge 1$, if there is a recursive dynamics that continues beyond the first 
redistribution step, then it initiates from the fibers having highest thresholds.
As mentioned in the main text, the increment of the load on a fiber with failure threshold $f_i$ in the first
redistribution step, following a sudden application of load $\sigma$ is
\begin{equation}
 x(f_i)=\frac{\sigma^2(b+1)}{(1-\sigma)^{b+1}}(f_i-\sigma)^b.
\end{equation}  
Hence the total load on a fiber with threshold $f_i$ becomes $\sigma+x(f_i)$. Since all fibers having threshold on or below $\sigma$
have already broken immediately after the application of load, all remaining fibers are of threshold higher than $\sigma$. Hence we can
write 
\begin{equation}
 f_i=\sigma+\epsilon_i.
\end{equation}

\begin{figure}[tbh]
\centering 
\captionsetup{justification=raggedright}
\includegraphics[width=8cm]{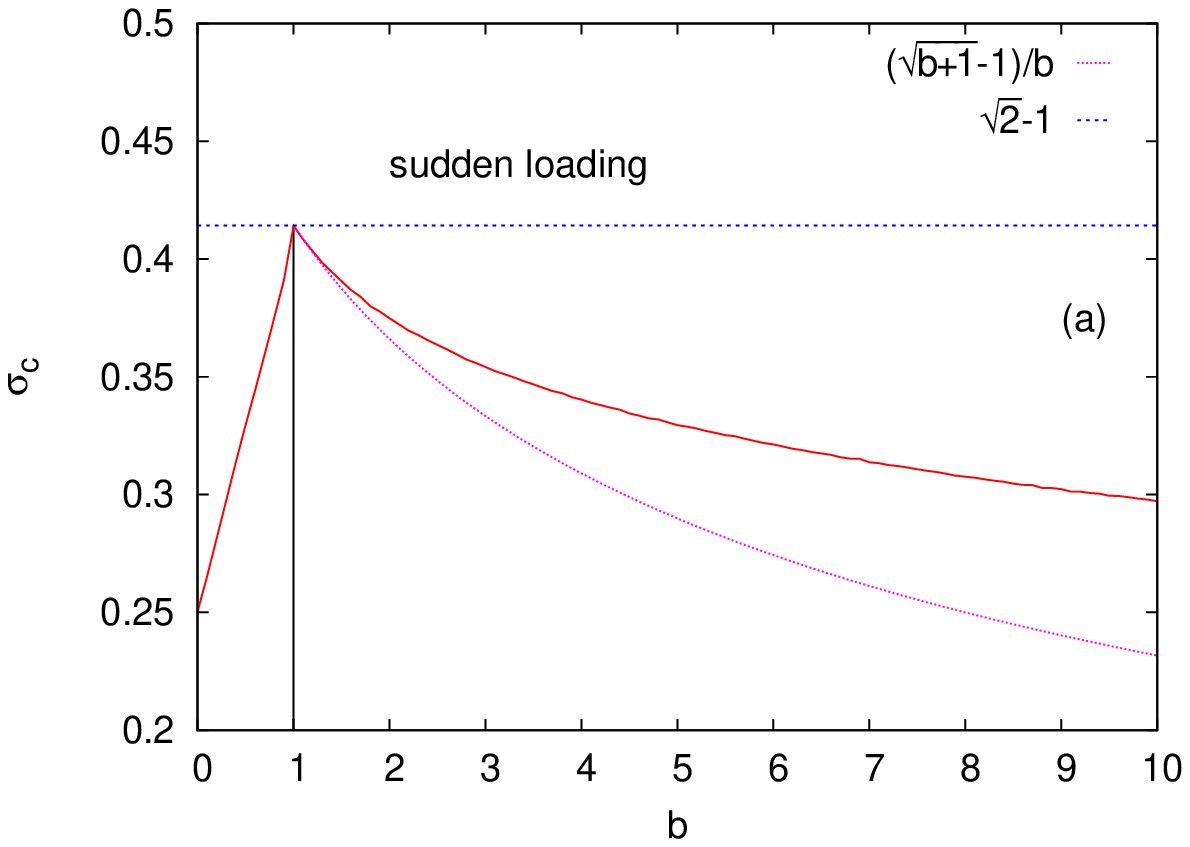}
 \includegraphics[width=8cm]{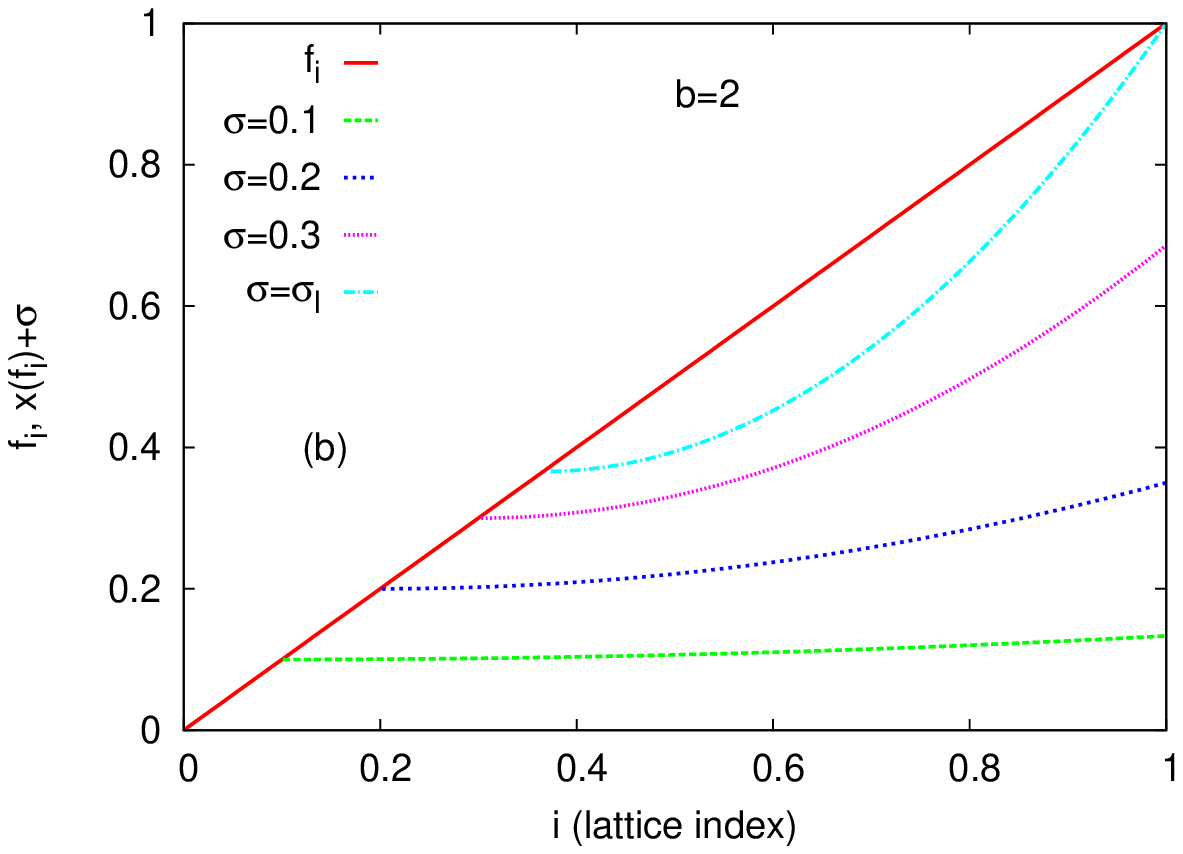}
   \caption{(a) The critical stress for sudden loading is plotted (this corresponds to the $e=0$ line in Fig. 1 (a)) 
along with the lower-bound (Eq. \ref{lower_bound}) for $b\ge 1$. 
The lower bound essentially marks the onset of recursive dynamics and hence matches with the critical load for $b=1$ where the process
is single step. (b) The stress profiles on the fibers (arranged in ascending order of their failure thresholds) after the first redistribution
step for different $\sigma$ values and for $b=2$. For $\sigma \ge \sigma_l$ ($\approx 0.366$ in this case), the fibers with high threshold values will break,
indication the onset of recursive dynamics.}
\label{op_lb}
\end{figure}
First we show that for very small $\epsilon_i$, the fiber survives. The condition for survival for a fiber is (dropping the index $i$)
\begin{eqnarray}
x(f)+\sigma &<& f \nonumber \\
\Rightarrow\frac{\sigma^2(b+1)}{(1-\sigma)^{b+1}}\epsilon^b+\sigma &<& \sigma+\epsilon \nonumber \\
\Rightarrow\frac{\sigma^2(b+1)}{(1-\sigma)^{b+1}}\epsilon^{b-1}&<& 1.
\end{eqnarray}
This is satisfied for small $\epsilon$ and when $b>1$ (and $\sigma$ not close to unity, which would in any case break the entire system).
Hence the breaking of the fibers will begin from the highest threshold value (unity in this case). 
This picture is also supported in Fig. \ref{op_lb} (b), where we have plotted the load profiles on the fibers (arranged in ascending order of their
threshold values) after the first redistribution step for different $\sigma$ values. As can be seen from the figure, there is a value $\sigma_l$, on
or above which more fibers will break in this step, marking the onset of recursive dynamics. The value of $\sigma_l$ can be estimated from the
condition that it is the load for which the fiber with highest threshold value (unity) will break after first redistribution. Hence
\begin{equation}
x(f_i=1)+\sigma_l=1.
\end{equation}
This gives,
\begin{equation} 
\sigma_l=\frac{\sqrt{1+b}-1}{b},
\label{lower_bound}
\end{equation} 
which is essentially a lower bound of $\sigma_c$ for $b>1$. These are also
supported by numerical simulations (see Fig. \ref{op_lb} (a)).
An estimate of this lower bound may be useful for safe designs.

\section{Calculation of the normalization constant A}
The normalization condition (Eq. (1) in the main text) reads
\begin{equation}
  \sum\limits_{i\in \mathcal{S}}\int\limits_{\sigma}^{1}\left[A(f_i-\sigma)^b\sigma^2N+\sigma \right]P(f_i)df_i=\sigma N,
\end{equation}
while the sum is performed on the surviving fibers (i.e. $i\in\mathcal{S}$) following an application of load $\sigma$ on each fiber (hence the limit
in the integral), the right hand side is the total applied load, which is to be conserved. Performing the integral
\begin{equation}
\sum\limits_{i\in\mathcal{S}}\left[A\left.\frac{(f_i-\sigma)^{b+1}}{b+1}\right|_{\sigma}^{1}\sigma^2N+\sigma(1-\sigma)\right]\frac{1}{1-\sigma}=\sigma N,
\end{equation}
where we have used $P(f)\propto 1/(1-\sigma)$ for the remaining fibers. Now the sum gives a factor $(1-\sigma)N$, which is the number of remaining fibers.
Putting that in the above equation we have 
\begin{equation}
A=\frac{(b+1)}{N(1-\sigma)^{b+1}},
\end{equation}
which is used in the main text.

\section{Calculation of upper bound of failure threshold for sudden loading}
In the main text, the upper bound of failure threshold ($\sigma_m$) was derived for the threshold distribution
uniform in $[0:1]$. Here we provide the general expression for arbitrary threshold distribution and then focus
on the specific type of uniform distribution (between $[a:1-a]$) for which the mean is fixed at $1/2$. The $a\to 0$
limit is the one we studied in the main text. $\sigma_m$ has non-monotonic variation with $a$.
 
\begin{figure}[tbh]
\centering 
\captionsetup{justification=raggedright}
\includegraphics[width=6.5cm]{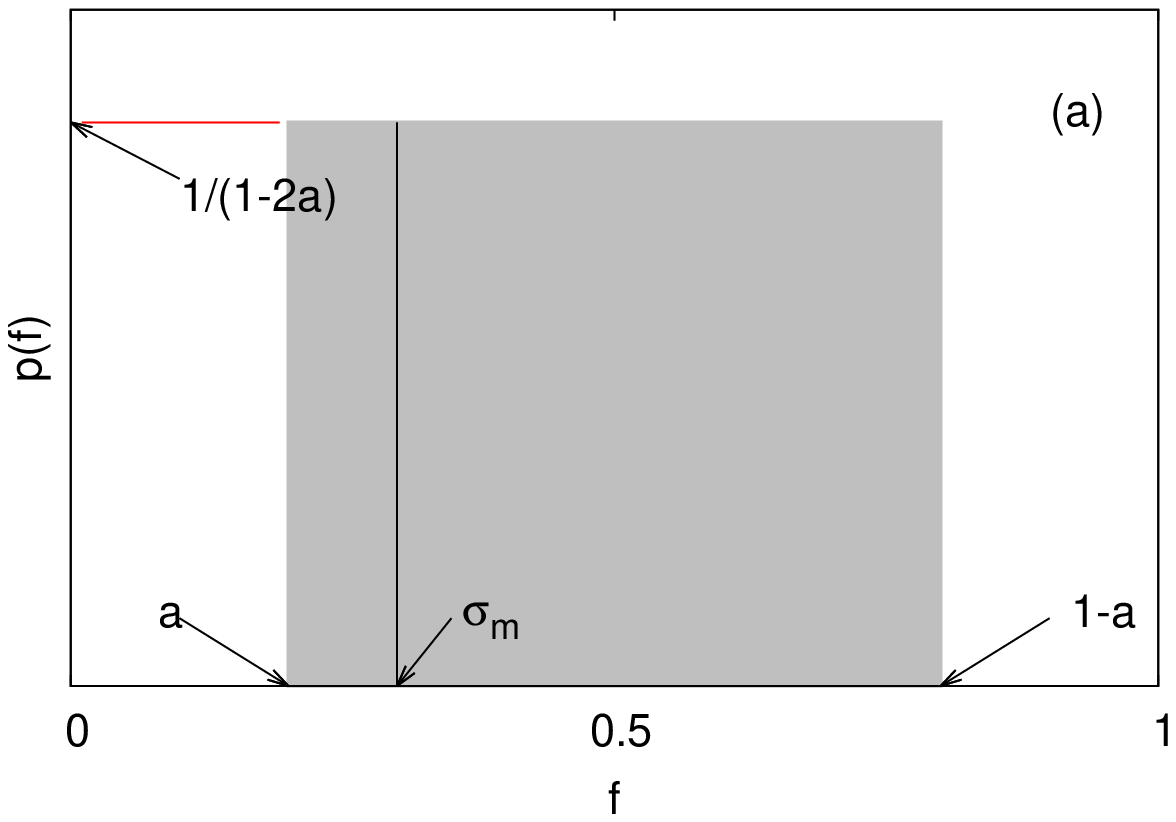}
 \includegraphics[width=7.5cm]{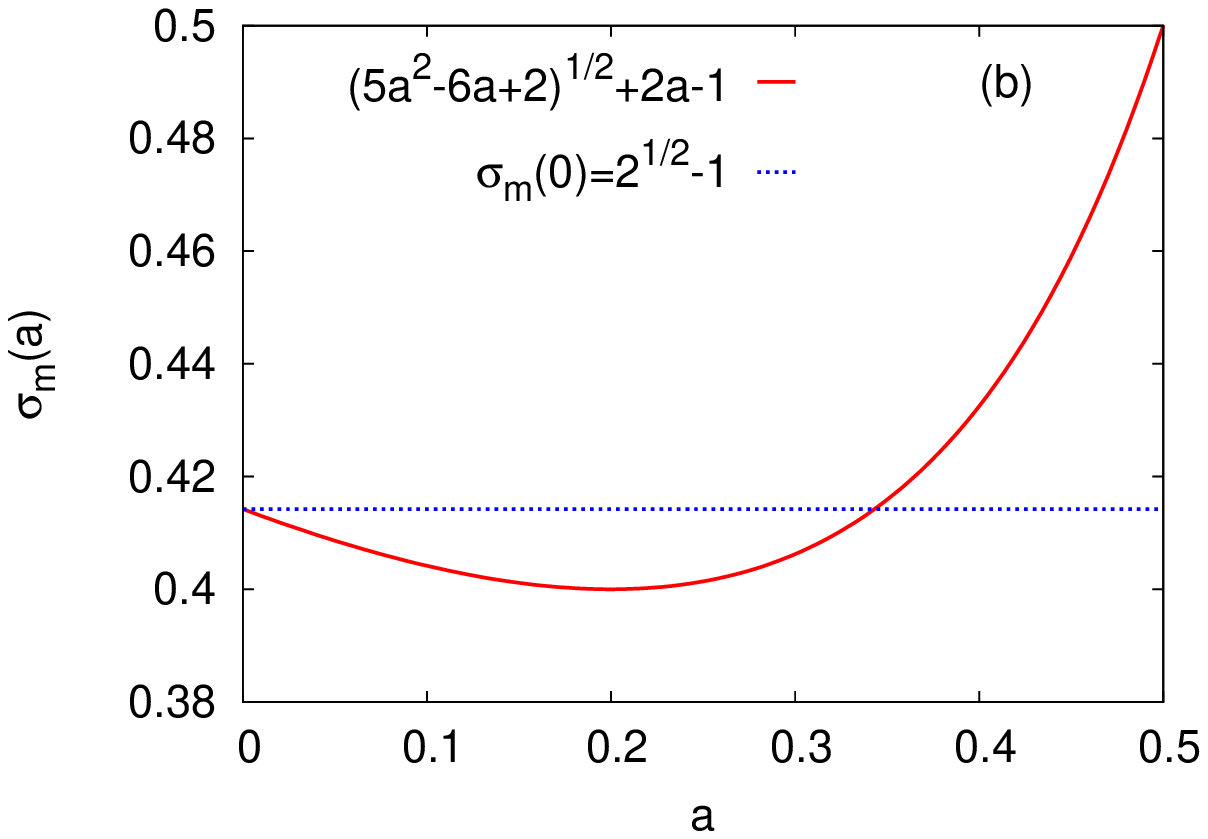}
   \caption{(a) The uniform threshold distribution in the range $[a:1-a]$ is shown with a schematic diagram. 
The maximum critical load $\sigma_m$ is also indicated. (b) The variation of $\sigma_m$ with the parameter $a$
is shown. The $a\to 0$ limit is discussed in the main text. The maximum value occurs at $a\to 1/2$ limit, which is the
limit of vanishing disorder (delta function limit). The non-monotonicity arises out of the competition of
less number of broken fiber on application of load $\sigma_m$ and the presence of smaller number of stronger 
fiber as $a$ is increased.}
\label{thr_dist}
\end{figure}
First consider an arbitrary threshold distribution function $p(f)$ 
defined in $[0:\infty]$. If a load $\sigma_m$ is applied uniformly
on all fibers, then all fibers having threshold values less than $\sigma_m$ will break immediately. 
Now, the best possible scenario (with regard to higher failure threshold) is the one where no further
fiber breaks and all of them carry load upto their maximum capacity. This means
\begin{equation}
\int\limits_{\sigma_m}^{\infty}fp(f)df=\sigma_m.
\label{max_sig}
\end{equation}
Solution of this equation will yield the maximum failure threshold for sudden loading 
$\sigma_m$ for an arbitrary threshold distribution.

Now let us focus on the uniform threshold distribution with its mean fixed at $1/2$ (see Fig. \ref{thr_dist} (a)). This
is a generalization of the particular distribution we considered in the text. The distribution has
the limits $[a:1-a]$. So, the normalized distribution is
\begin{equation}
p(f)=\frac{1}{1-2a}.
\end{equation}
Applying Eq. (\ref{max_sig}) we have 
\begin{equation}
\frac{1-a-\sigma_m}{1-2a}\frac{1-a+\sigma_m}{2}=\sigma_m,
\end{equation}
of course the upper limit of the integral here is $1-a$.
Solving it one has
\begin{equation} 
\sigma_m=\sqrt{5a^2-6a+2}+2a-1.
\end{equation}
The $a\to 0$ limit gives $\sigma_m(0)=\sqrt{2}-1$, as discussed in the main text. The plot of $\sigma_m$ versus $a$
is shown in Fig. \ref{thr_dist} (b). In the $a\to 1/2$ limit, the threshold distribution becomes a delta function
centred at $1/2$. That is of course the highest value of $\sigma_m$. But before that it shows a
non-monotonicity with $a$. This is because, increasing $a$ brings a competition between 
less number of broken fibers with less number of strong fibers present.

\end{document}